# Compute-in-Memory Implementation of State Space Models for Event Sequence Processing

Xiaoyu Zhang[1,§], Mingtao Hu[1,§], Sen Lu[1], Soohyeon Kim[1], Eric Yeu-Jer Lee[1], Yuyang Liu[1] and Wei D. Lu[1,*]

[1]Department of Electrical Engineering and Computer Science, University of Michigan, Ann Arbor, MI 48105, USA.

[§]These authors contributed equally.

* Correspondence to: wluee@umich.edu.

## Abstract

State space models have recently emerged as a powerful framework for long sequence processing, outperforming traditional methods on diverse benchmarks. Fundamentally, state space models can generalize both recurrent and convolutional networks and have been shown to even capture key functions of biological systems. Here we report an approach to implement SSMs in energy-efficient compute-in-memory hardware to achieve real-time, event-driven processing. Our work re-parameterizes the model to function with real-valued coefficients and shared decay constants, reducing the complexity of model mapping onto practical hardware systems. By leveraging device dynamics and diagonalized state transition parameters, the state evolution can be natively implemented in crossbar-based compute-in-memory systems combined with memristors exhibiting short-term memory effects. Through this algorithm and hardware co-design, we show the proposed system offers both high accuracy and high energy efficiency while supporting fully asynchronous processing for event-based vision and audio tasks.

## Introduction

Sequence modeling is a fundamental computational task essential for diverse fields, from natural language processing[1] to time-series data analysis[2]. A significant focus within the area of neuromorphic computing lies on sequential information processing, particularly efficiently processing asynchronous, event-based data streams. Such data can originate directly from sensors such as event cameras, which capture changes in luminance and generate sparse events[3,4], or be derived by converting traditional datasets into spike-based formats[5] to take advantage of sparsity in communication and processing.

Spiking Neural Networks (SNNs) are generally the model of choice for event data processing, as their fundamental operation is based on asynchronous spike events, mirroring the data format[6]. The asynchronous, event-driven nature holds considerable potential for energy-efficient data processing, as computation is ideally performed only when new data arrives[7,8].

However, despite their potential, training deep SNNs can be challenging due to the non-differentiable nature of spike generation, which necessitates the use of surrogate gradient techniques for backpropagation[9]. While these techniques have proven increasingly robust[10,11], the inherent binarization of activations may still limit model performance, a constraint recently identified as a key factor for scaling up SNNs[12]. Tradeoffs between different encoding methods and neuron models have also been found to affect model performance[13].

To improve model accuracy, SNNs often incorporate spatial processing blocks such as convolution or transformer blocks to help extract spatial features[14–16]. However, the use of these blocks requires collecting a sequence of spikes and converting the sequence of asynchronous events into a frame-based representation, thereby sacrificing the inherent sparsity and low-latency advantages of the event-based approach. This mismatch between asynchronous data inputs and synchronous processing algorithms motivates the search for more suitable models.

Furthermore, from a hardware perspective, convolution and transformer blocks used in the SNN models are not well suited for energy efficient architectures such as compute-in-memory (CIM). In CIM, weights are stored stationarily in high-density memory cells such as Resistive RAM (RRAM)[17–19] or Phase-Change Memory (PCM)[20,21], and computations, primarily vector-matrix multiplication (VMM), are directly performed inside the memory array through Ohm's law and Kirchhoff's current law[22,23]. This in-situ computation can dramatically minimize data movement compared to traditional von Neumann designs[24–26]. However, for convolution operations, the same weights are re-used for many input patches thus the weight read costs can be efficiently amortized in well-designed digital systems, diminishing the benefits of CIM and amplifying the challenges of the slower read in emerging devices used for CIM implementations. The need to load different input patches from the feature map memory to perform convolutions also requires data re-shuffling, again diminishing CIM benefits. As a result, convolution operations are generally not well matched with CIM implementations. In Transformers, the attention mechanism involves the multiplications of the Query, Key, and Value matrices, all of which depend on the input so the attention operation is not friendly to stationary CIM arrays which assume one of the matrices (i.e. weights) is static. Consequently, practical implementations of Transformers adopt hybrid designs: they leverage stationary CIM only for VMM involving static weights (e.g., in Fully Connected Layers), but delegate the input-dependent attention score computations or the KV cache management to separate digital circuitry[27].

To overcome both the architectural mismatch with convolutions and the memory and scaling bottleneck inherent in Transformers, in this work, we present a State Space Model (SSM) based approach that is natively compatible with CIM hardware. We show that the proposed SSM allows asynchronous spike sequence processing with high model performance without the need for explicit spatial processing blocks (*e.g.* convolutions). SSMs have recently gained prominence for sequence modeling especially on tasks requiring the capture of long-range dependencies[28,29], offering strong performance compatible or exceeding that of recurrent, convolutional, and Transformer-based methods. Grounded in control theory with theoretical guarantee to stably approximate continuous-time convolutional kernels, SSMs model the sequence through a latent state vector that evolves over time based on the current input and the previous state[30]. A key innovation in recent SSMs like S4[31] and S5[32] is the use of a diagonalized state transition matrix, which allows for highly parallelizable computation of the state updates and efficient modeling of long-term dependencies through exponential decay dynamics. S4 sets new state-of-the-art on the Long-Range Arena—solving the 16 384-length Path-X task with 91 % accuracy and running 60× faster than Transformer-based models. Its successor S5 matches S4's efficiency while averaging 87.4 % on LRA and 98.5 % on Path-X. Subsequent developments such as Mamba[33] and Mamba2[34] employ selective SSM layers to achieve linear-time scaling to million-length contexts.

Beyond superior model performance, SSMs have been shown to generalize recurrent, convolution, and attention-based models[34,35], offering a unified theoretical framework for sequential data processing. SSMs do not rely on sliding-window operations such as convolutions or Transformer blocks, suggesting they are compatible with asynchronous spike processing[36] and also potentially more friendly to CIM hardware. However, mapping SSMs onto practical CIM hardware presents its own challenges. Firstly, many high-performing SSMs (e.g., S4, S5) rely on complex-valued state representations and computations. Emulating complex arithmetic using real-valued circuits such as CIM introduces overhead in terms of complexity, area, or power. Secondly, SSMs inherently rely on maintaining and updating a hidden state that captures temporal context, often involving specific dynamics. While CIM excels at VMM, relying on external digital components (e.g., registers, digital logic) to store and update the state can reintroduce significant overhead between the analog CIM core and digital peripheral, negating energy and latency benefits.

In this study we address these challenges through a co-design approach, by developing both a hardware-friendly SSM algorithm and a hardware architecture for efficient SSM mapping and execution on event-based data. While recent works have successfully mapped SSMs onto digital neuromorphic processors[37,38], our approach represents a co-design methodology leveraging analog CIM to perform key operations, including state update, directly through device physics with potentially greater efficiency and scalability. We first re-parameterize the SSM by restricting the state transition dynamics to operate purely with real-valued coefficients, eliminating the overhead associated with complex arithmetic in hardware. Furthermore, we constrain the exponential decay parameter ($\lambda$) associated with the state dynamics to one (or a few) fixed value(s) per block, significantly simplifying hardware implementation. Subsequently, we co-design the hardware architecture to natively implement all major operations, including the state dynamics, in analog domain through physics. The hardware architecture consists of modular, asynchronous blocks that match the functions of SSM blocks. Within each hardware block are RRAM crossbar arrays that perform the necessary VMM operations, and a group of memristors exhibiting short-term memory (STM) effects (e.g. Tungsten Oxide ($WO_x$) based memristors[39]) to natively store the state components and implement the diagonalized state update functions. The physical decay rates ($\lambda$) of these STM devices are tuned during fabrication to align with the few fixed values required by the modified SSM algorithm. Overall, our co-designed system using real-valued SSM with fixed decays consistently matches or exceeds the performance of leading models on both audio and vision event-stream benchmarks with small number of parameters.

By leveraging a core asynchronous state evolution mechanism based on device physics, the proposed hardware implementation offers an efficient pathway for event-driven, real-time processing. These results suggest a promising approach for accurate, scalable, and highly-efficient neuromorphic systems capable of processing real-time event streams.

# Results

## Model Architecture

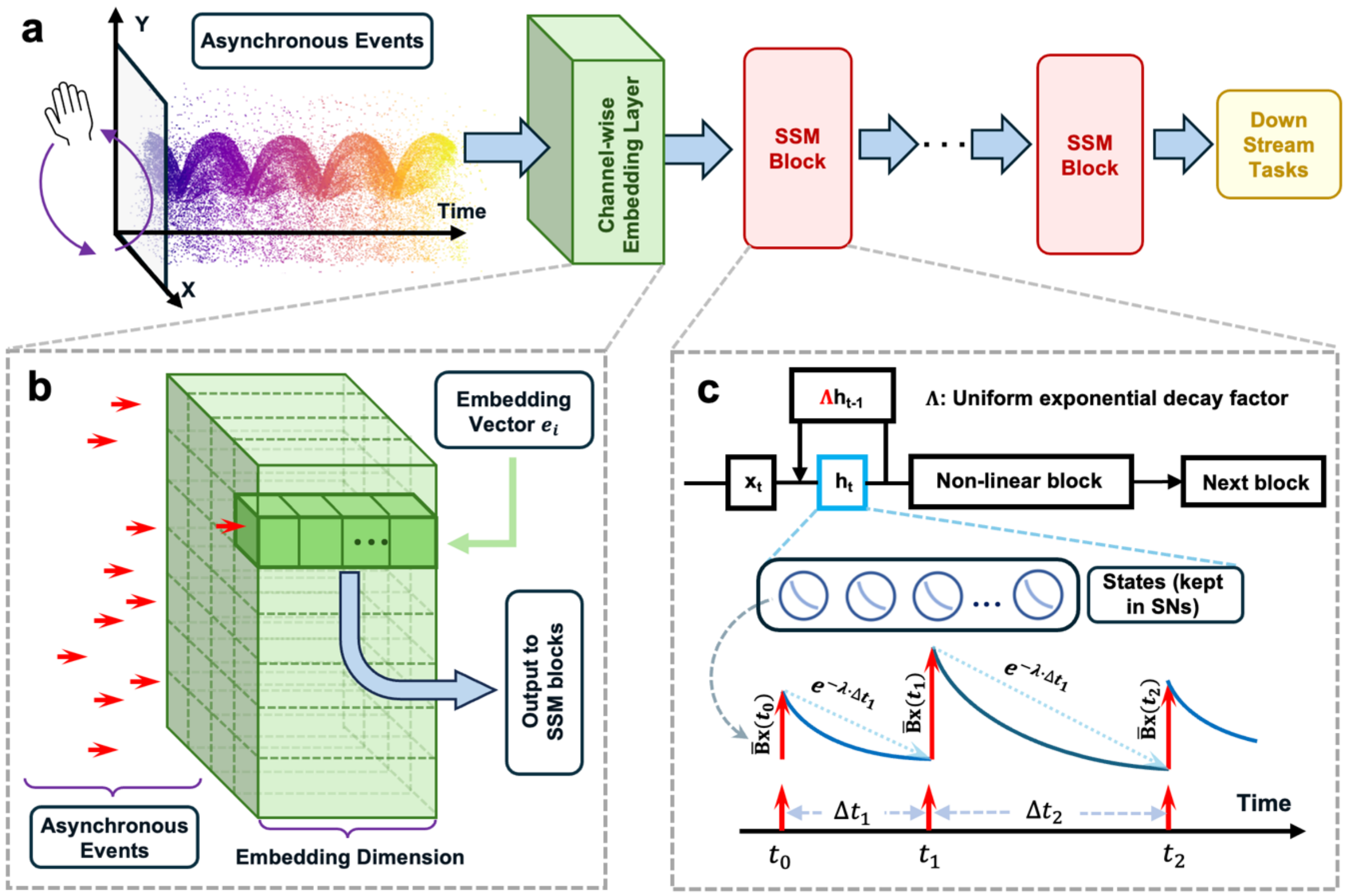

Fig. 1a shows the high-level model architecture, which is designed for asynchronous event processing. The input event-stream is an ordered set of events $E = \{(t_m, j_m)\}$, where $t_m$ is the timestamp and $j_m$ is the channel index of event $m$. In audio tasks, channel indices correspond to distinct frequency bands whereas in vision tasks, channels map one-to-one with pixel locations. The channel-wise embedding layer, shown in Fig. 1b, maps an event into a D-dimensional vector based on the event's channel index. Since the embedding matrix $W_{\text{embedding}} \in \mathbb{R}^{J \times D}$, where $J$ is the number of input channels, is static after training, it can be physically stored in an RRAM crossbar array and embedding is performed by reading out the values in the row corresponding to $j_m$ when spike $m$ arrives. Embedding is a common pre-processing technique used in modern models, e.g. language models. In this case, by mapping each channel to a distinct embedding vector, we endow every channel with a unique feature representation, enabling the SSM blocks to process event streams within a unified feature space.

The embedded event representations are then passed through a series of stacked SSM blocks. Fig. 1c shows the internal operations of a single SSM block. To simplify hardware implementation, we modify the complex-valued decompositions in the original SSM models to perform all operator factorizations in real domain (see **Methods**). Specifically, the continuous-time dynamics of the state are modeled by a first-order differential equation:

$$\frac{\mathrm{d}\mathbf{h}(t)}{\mathrm{d}t} = \mathbf{\Lambda}\mathbf{h}(t) + \mathbf{B}\mathbf{x}(t), \tag{1}$$

where $\mathbf{h}(t) \in \mathbb{R}^{H}$ denotes the hidden state, $\mathbf{x}(t) \in \mathbb{R}^{H_{\text{in}}}$ is the input, $\mathbf{\Lambda} \in \mathbb{R}^{H \times H}$ is a diagonal matrix encoding the state transition rates, and $\mathbf{B} \in \mathbb{R}^{H \times H_{\text{in}}}$ is the input projection matrix. An output matrix $\mathbf{C} \in R^{H \times H_{\text{out}}}$ will be used to convert the hidden state into output vector for downstream tasks or to the next layer.

To implement the model in an event-driven framework, we apply an asynchronous discretization method following Schöne et al.[36]. For a time interval $\Delta t$ between two spike events at $t$ and $t + \Delta t$, the state update equation becomes

$$\mathbf{h}(t + \Delta t) = \overline{\mathbf{\Lambda}}(\Delta t)\mathbf{h}(t) + \overline{\mathbf{B}}\mathbf{x}(t + \Delta t), \tag{2}$$

where the state transition matrix $\overline{\mathbf{\Lambda}}(\Delta t)$ and the input projection matrix $\overline{\mathbf{B}}$ are defined as

$$\overline{\mathbf{\Lambda}}(\Delta t) = \exp(\mathbf{\Lambda}\,\Delta t)\,, \tag{3}$$

$$\overline{\mathbf{B}} = \mathbf{\Lambda}^{-1}(\exp\,(\mathbf{\Lambda})\ \ - \mathbf{I})\mathbf{B} \tag{4}$$

where $\mathbf{I}$ is the identity matrix. We note that Eq. (2) can then be efficiently implemented in analog circuits where the 2nd term corresponds to a VMM function and can be implemented using an RRAM crossbar for the current input at $t + \Delta t$, while the 1st term can be implemented in parallel using a group of devices (termed state nodes, SNs here) exhibiting decay dynamics, taking advantage of the diagonalized matrix $\mathbf{\Lambda}$.

As illustrated in Fig. 1c, When no inputs are applied (e.g., during intervals $\Delta t_1$ and $\Delta t_2$), the state evolve passively and natively through the short-term memory property of the SNs, implementing the exponential decay (e.g., $e^{-\Delta t_1}, e^{-\Delta t_2}$ ) without the need of clocks or synchronization. Active computation is needed only when an input arrives (e.g., at time $t_1$, $t_2$), where the projected input term (e.g., $\overline{\mathbf{B}}\mathbf{x}(t_0)$, $\overline{\mathbf{B}}\mathbf{x}(t_1)$, $\overline{\mathbf{B}}\mathbf{x}(t_2)$) is computed via VMM and added to the current state, making the system operation fully event-driven. The updated state is then read out, further processed through a non-linear block, and applied as an input event to trigger the next block's computation. By leveraging physics, including Ohm's law, Kirchhoff's current law, and internal device dynamics, this approach achieves highly parallel and energy-efficient computation and is compatible with asynchronous inputs.

To further facilitate practical implementation, a key design choice in our architecture is to limit one SSM Block to the same decay rate in the matrix $\mathbf{\Lambda}$. In other words, instead of learning independent decay rates for each dimension in the state, we set

$$\mathbf{\Lambda} = \lambda\,\mathbf{I}, \tag{5}$$

where $\lambda$ is a hyperparameter determined during training (Supplementary Notes 1). At first glance, imposing a single shared decay rate across an entire block may seem excessively restrictive. However, in practice, we found that this constraint does not notably limit the model's representational capacity. This is likely because other learnable parameters, such as the input projection matrix **B** and output matrix **C**, along weights within the nonlinear block, can compensate the restrictions on $\lambda$. Furthermore, the stacking of multiple blocks provides additional mechanisms for modeling diverse temporal dynamics. If necessary, the constraint can be relaxed by grouping state variables into a few clusters within one block, each with its own decay constant - *i.e.* having *n* (where $n \ll H$ which is the dimension of the state space) per block. This adjustment can better balance the requirements for practical hardware implementation and preserving sufficient expressive power for the model, although for the datasets we tested having a single time constant per block already offers excellent model performance.

Interestingly, Eq. 2 closely resembles the dynamics of a Leaky Integrate (LI) neuron, in which the neuron's membrane potential decays exponentially over time. This similarity suggests that SSMs could be biologically feasible, providing an intriguing connection to neural mechanisms observed in nature. However, a key benefit of SSM is that most operations are linear (e.g. Eq. (1) only has linear terms), avoiding strong non-linearity such as spike generation and reset operations at the neuron level which is a hallmark for SNNs. This property makes it highly effective to train SSMs, as discussed in Orvieto et al.[40]. The only positional non-linear function is at the final output of an SSM block, in the form of a gated activation function such as:

$$\tilde{\mathbf{h}} = \mathbf{h} \odot \sigma(\mathbf{W} \cdot \mathrm{GELU}(\mathbf{h}) + \mathbf{b}), \tag{6}$$

where $\mathrm{GELU}(\cdot)$ is the Gaussian Error Linear Unit activation, $\mathbf{W}$ and $\mathbf{b}$ are learned parameters, $\sigma(\cdot)$ denotes the sigmoid function, and $\odot$ represents element-wise multiplication. $\tilde{\mathbf{h}}$ is then combined with the original state via a residual connection:

$$\mathbf{h}_{\mathrm{out}} = \mathbf{h} + \tilde{\mathbf{h}}. \tag{7}$$

## Hardware Implementation of State Space Model Block

Having detailed the mathematical framework of the proposed SSM, we now describe its physical implementation in a CIM-based hardware. We co-designed the hardware architecture to natively execute the model's core operations, following the principles of performing computing directly through device and circuit physics. Specifically, we implement the state update equation by leveraging the native exponential decay of the internal state variable in memristor devices exhibiting STM effects[39]. For example, prior studies have shown that when stimulated by an input, the $WO_x$-based memristor device conductance increases according to the input then spontaneously relaxes due to oxygen ion diffusion, resulting in STM behavior[41,42]. Given that state parameters evolve independently (due to the use of diagonal state dynamics matrix) in an SSM block, the state can be represented by an array of such vector decay dynamics can be parallelized and directly implemented using STM memristors (SNs).

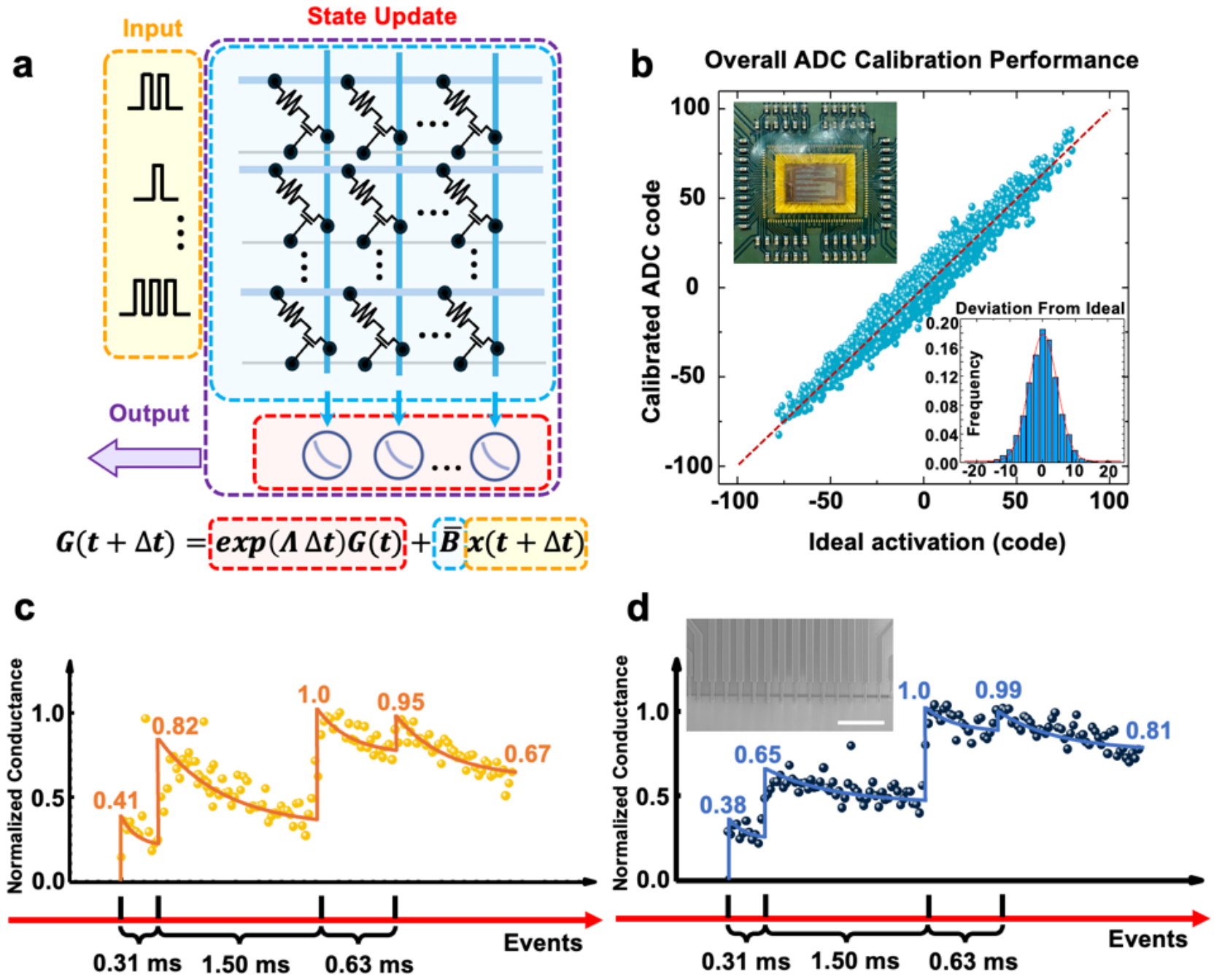

Specifically, the state update equation inside an SSM block (Eq. (2)) can be implemented as:

$$\mathbf{G}(t+\Delta t)=\exp(\mathbf{\Lambda}\,\Delta t)\mathbf{G}(t)+\overline{\mathbf{B}}\mathbf{x}(t+\Delta t). \quad (8)$$

Here $\mathbf{G}(t)=[g_0(t),\ g_1(t)\ldots\ g_h(t)]$ is the collection of the STM memristor's conductance at time $t$ and represents the hidden state vector, $\mathbf{\Lambda}=\lambda\,\mathbf{I}$ and $\lambda$ is the layer-wise fixed decay constant, $\Delta t$ is the time interval since the previous event, $\overline{B}$ is the input projection matrix and $\mathbf{x}(t+\Delta t)$ is the input at time $t+\Delta t$ when the current spike arrives. We note that this equation can be physically implemented using an array of STM memristors that act as SNs to store the state and implement the state decay dynamics ($1^{st}$ term in Eq. (8)), and an RRAM crossbar that implements the VMM operation between the $\overline{B}$ matrix and the input ($2^{nd}$ term in Eq. (8)), as shown in Fig. 2a. The proposed hardware thus directly implements the state update equation (Eq. 8), which in turn corresponds to the state equation Eq. 2.

We tested this implementation experimentally using an RRAM CIM chip that performs the VMM operations and a group of $WO_x$ memristors that exhibit STM behaviors, respectively. The inset of Fig. 2b shows a photograph of the RRAM CIM chip for the VMM operations. This chip was fabricated using a 65 nm CMOS process with four integrated 64×64 RRAM arrays, which are logically combined to form an equivalent 128×128 array. Each RRAM device in the array is based on a one-transistor-one-resistor (1T1R) structure with a physical size of 200 nm[43]. Each array is connected with integrated 8-bit digital-to-analog converters (DACs) for input encoding and 8-bit analog-to-digital converters (ADCs) for output readout. The measured VMM outputs from the RRAM CIM chip are shown in Fig. 2b (dots), along with the ideal expected values (line).

The measured output distribution (bottom inset, Fig. 2b) due to device variations is characterized by a standard deviation $\sigma$ of 4.6 LSB (Least Significant Bit), corresponding to a Normalized Root-Mean-Square Error (NRMSE) of 1.80%. The high consistency between measured and ideal values and low NRMSE verify that the RRAM CIM configurations can accurately performing the VMM operation required for the $\overline{\mathbf{B}}\mathbf{x}(t+\Delta t)$ term in Eq. 8. More details of the RRAM CIM chip can be found in Supplementary Figure 1 and Supplementary Note 2.

The VMM output is then applied to the SNs to implement the exponential state evolution ($1^{st}$ term in Eq. (8)). Fig. 2c-d shows the state evolution function implemented with $WO_x$ memristors that exhibit STM behaviors, where the $\exp(\mathbf{\Lambda}\,\Delta t)\mathbf{G}(t)$ function is passively implemented based on the native conductance decay in between inputs. The application of the VMM output updates the decayed conductance values, completing the state transition by naturally summing the two terms of Eq. 8 in the STM memristor devices.

To match the required decay profiles, $WO_x$ devices with different oxide thicknesses were fabricated by tuning the rapid thermal processing (RTP) annealing duration (see **Methods**), following prior studies[41,44]. Longer annealing times yield thicker oxide layers and higher oxygen vacancy densities, which in turn slow the relaxation process due to more persistent conductive filament paths. More information on the tunable decay behavior can be found in Supplementary Figure 2.

Fig. 2c and 2d show experimentally implemented Eq. 8 with two target time constants. Strong agreement between measured outputs (scatter points) and the expected SSM-based evolution (solid lines obtained by simulation of Eq. (2)), confirming the viability of this approach for hardware SSM inference.

By fixing the decay constant to a single value in a block, only six distinct decay profiles—one per layer—are required for hardware implementation of the proposed SSM model. This significantly reduces the physical implementation challenges as precise control of the time constant τ to a different target value for every individual memristor device is impractical during fabrication.

The updated state from each SN is then read out and used to perform the positional non-linear gating and residual connection operations as described in Eq. 6 and Eq. 7. These operations, including the GELU and sigmoid activation functions, are implemented digitally using Look-Up Tables (LUTs) that store pre-computed values. The output is then passed to the next SSM block as a discrete input. This entire process, from event reception to SN state update, demonstrates the tight coupling between the mathematical formulation and physical device dynamics, enabling a highly efficient and clockless implementation of the SSM.

## Model Training and Performance Analysis

We now detail the training procedure used to determine the model's optimal parameters. We note that the training process is conducted in simulation. All standard network parameters—including the SSM input projection $\mathbf{B}$, the output projection matrix $\mathbf{C}$, the gating weights and biases, and (initially) each individual decay rate $\lambda$ inside each layer—are trained jointly with backpropagation through time (BPTT)[45], without any surrogate approximations (see **Methods**). To reconcile the need for per-dimension flexibility during learning with the hardware constraint of a single shared $\lambda$ per block, we determine $\lambda$ in three stages. First, we train the entire network (including each $\lambda$) until the training loss plateaus—allowing each state dimension to discover its own decay rate. Second, we compute the arithmetic mean of all learned $\lambda$ within each layer. Finally, we fix $\lambda$ to this layer-wise average and continue training the remaining parameters for all subsequent epochs. The detailed information about the $\lambda$ for each task is shown in Supplementary Figure 3 and Supplementary Table 1.

We first trained and evaluated model on four event-driven datasets spanning both spiking audio and event-based vision tasks. The spiking audio datasets include Spiking Heidelberg Digits (SHD)[5] and Spiking Speech Commands (SSC)[5]. SHD comprises spoken digits converted into spike trains through a biologically inspired auditory model, with each digit represented as a spatio-temporal pattern across multiple frequency channels, while SSC is derived from Google's Speech Commands corpus, retaining word categories like "yes," "no," "up," and "down," but converting them into spike trains via a gammatone-like filter bank for denser and lengthier event streams. The event-based vision datasets are DVS128 Gesture[3] and DVS128 Lips[4], both captured by a DVS camera, producing asynchronous spike sequences on a 128×128 pixel grid. DVS128 Gesture encompasses diverse hand and arm movements that demand modeling both the temporal progression and limited spatial information of the gesture pattern, whereas DVS128 Lips highlights subtler lip movements often used for audio-visual speech processing, requiring fine-grained temporal encoding to distinguish rapid motion cues. To improve robustness and generalization under real-world use cases, we applied standard data augmentations during training[46]. The training loss curves are shown Supplementary Figure 4.

Table 1 summarizes the simulation results on two prominent spiking audio datasets, which are frequently tackled with SNN-based architectures. As shown in Table 1, our real-valued asynchronous SSM model with shared decay time constants consistently outperforms SNN-based baselines while maintaining a low parameter count. Notably, even though our method does not explicitly replicate the neuron-level spiking mechanism found in SNNs, it still excels at handling these sparse, event-driven audio signals. The results demonstrate that SSMs with diagonalized exponential-decay dynamics is both robust and effective for long-range temporal tasks consisting of asynchronous spiking sequences.

Results on two representative event-based vision datasets—DVS128 Gesture and DVS128 Lips—as shown in Table 2. Unlike many existing works on DVS data, we do not include commonly deployed spatial modules such as convolutional operations, since these operations 1) do not support asynchronous processing (as a full feature map needs to be created first to support the scanning of input patches) and 2) are not friendly with CIM hardware, as discussed earlier. Instead, our model relies on purely temporal SSM blocks, where the spatial features are encoded through the embedding process and the attention-like state updates.[34] Despite this choice of avoiding convolution operations, our approach achieves accuracy on DVS128 Gesture that generally matches more complex architectures integrating explicit spatial processing. On DVS128 Lips, we observe similarly competitive performance with a parameter count significantly lower than that of leading methods.

These results underscore that an event-driven SSM framework—even with simplified real-valued dynamics—can generalize across vision tasks on sparse, asynchronous data without the overhead of spatial convolutions. Equally importantly, the asynchronous nature of our SSM implementation can fully take advantage of the spiking inputs and the parallel computing capabilities of CIM systems for highly efficient hardware implementation.

It is noteworthy that some synchronous methods, by aggregating sparse events into frame-like structures, can achieve higher accuracy on DVS dataset. However, this performance comes at the cost of forgoing the benefits of event-based sensing. The frame-creation step not only introduces additional latency but also increases the data volume, leading to more computational demands. In contrast, the proposed asynchronous SSM processes each event as it arrives, preserving the data's inherent sparsity and temporal precision. This makes our approach suited for real-time, resource-constrained applications where minimizing latency and power consumption is paramount.

To provide a quantitative basis for the hardware and latency advantages of our approach, we performed a detailed computational complexity analysis (see **Methods**). For the DVS128 Gesture dataset, our Event-SSM requires 1.68 GFLOPs to process a typical data sequence. In contrast, a common alternative approach involves converting the event stream into video frames (e.g., at 30fps) and using a standard CNN. A ResNet-18 based model processing these frames would require 104.28 GFLOPs (a 62-fold increase), and a ResNet-50 would require 219.80 GFLOPs (a 131-fold increase). This significant reduction in computational load stems from our model's ability to leverage the inherent sparsity of input event data, avoiding the massive computational overhead associated with frame-based processing.

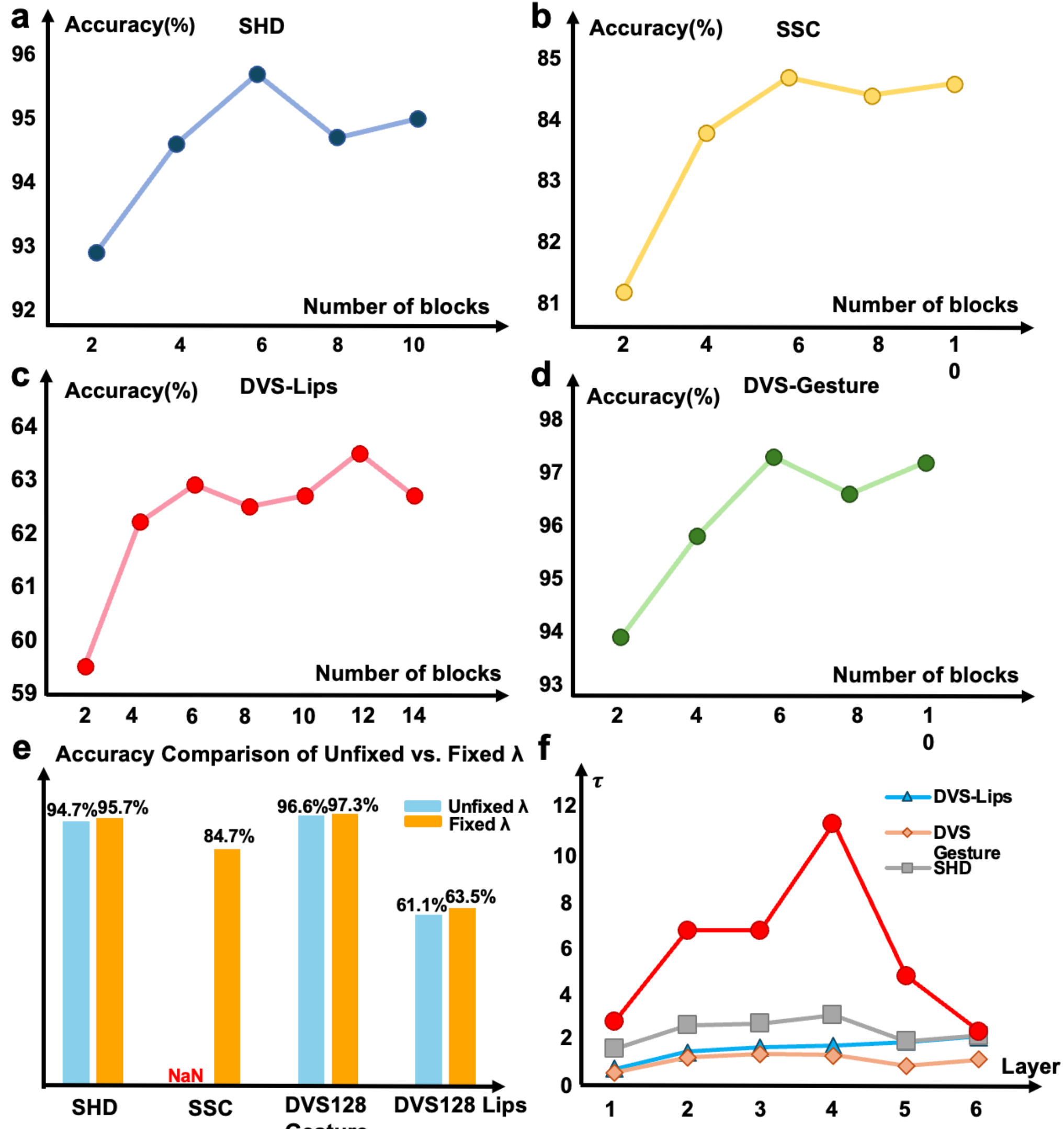

We performed several additional analyses in simulation to analyze the model behavior. Fig. 3 a-d shows how the classification accuracy evolves as we vary the number of stacked SSM blocks (model depth) for different event-driven datasets. All models tested employ the same underlying architecture differing mainly in the number of stacked SSM blocks after the embedding layer. We observe that increasing the depth initially enhances performance, reflecting the model's ability to better capture complex temporal dynamics with more layers. However, this improvement reaches a saturation point, typically around 6 blocks, indicating that adding further SSM blocks beyond a certain threshold yields minimal or no accuracy gain. The saturation effect is common in neural networks - beyond an optimal depth, adding more layers may offer diminishing returns and can increase the risk of overfitting or optimization difficulties during training. Therefore, the design of our model architecture is based on the saturation point. We note the required depth (about 6 blocks) is remarkably shallow compared to many conventional deep networks. For example, Bulzomi et al.[54] and Gehrig et al.[61] use ResNet[63] as the model backbone for the DVS datasets which employes more than 30 layers. The fact that SSMs can effectively model long-range event sequences with just 6 blocks underscores their representational efficiency and makes them particularly attractive for hardware deployment, where model mapping complexity and latency scale roughly with layer count.

To evaluate the effects of constraining the decay parameter to a single value per block, we compare the final classification accuracies across four event-driven datasets - two audio datasets and two vision-based datasets under conditions where the decay parameter $\lambda$ is either freely learned (unfixed) or held constant (fixed), shown in Fig. 3e. For most datasets, we observe that fixing $\lambda$ following the approach discussed in the Model Training section surprisingly results in improved accuracy by approximately 1% on average. Specifically, SHD improves from 94.7% to 95.7%, DVS128 Gesture from 96.6% to 97.3%, and DVS128 Lips from 61.1% to 63.5%. Notably, the Spiking Speech Commands (SSC) dataset exhibits a critical failure when $\lambda$ is unfixed, leading to divergent training with invalid Not a Number (NAN) loss values, yet it achieves a stable and competitive accuracy of 84.7% when $\lambda$ is fixed. These observations indicate that constraining $\lambda$ may in fact help stabilize training convergence by serving as a form of regularization for event-driven datasets, including those with challenging temporal structures such as SSC.

To further investigate the instability observed in SSC, we plot in Fig. 3f the distribution of $\tau = \frac{1}{\lambda}$ across the SSM layers for SSC and other datasets. Since training with unconstrained $\lambda$ on the SSC dataset terminated prematurely due to NAN values, the $\tau$ values plotted here for SSC were recorded from the final stable iteration immediately before divergence occurred. Notably, on SSC $\tau$ values are significantly larger compared to the other datasets, indicating slower decay rates in the hidden states. From a continuous-time perspective, the hidden state updates following:

$$\mathbf{h}(t + \Delta t) = \exp\left(-\frac{\Delta t}{\tau}\right)\mathbf{h}(t), \tag{9}$$

With larger $\tau$, the hidden state $h$ decays more slowly, preserving more historical information over prolonged periods. While beneficial for capturing long-term dependencies, this slow decay can cause the hidden states to accumulate excessively when the input event rate is high, which is also the case for SSC. Consequently, gradients and loss values can become unstable, resulting in divergence and NAN values. This numerical instability illustrates the challenges associated with unconstrained optimization of $\lambda$ on datasets with demanding temporal characteristics, and the necessity of carefully regularizing $\lambda$ to maintain numerical stability and achieve convergence.

## Hardware System Performance Analysis

We next evaluate the model implementation on CIM hardware using the SSC and DVS-Gesture datasets. To align the SSM model with analog CIM hardware constraints, all weights and activations are quantized to 8-bit integer (INT8) using quantization-aware training (QAT)[64,65]. This step causes no accuracy degradation for the DVS-Gesture model (remaining at 97.3%) and only a minimal 0.3% drop for the SSC model (from 84.7% to 84.4%).

We then studied the effect of device non-idealities in analog hardware by introducing random hardware variations to the INT8 parameters (see **Method**) based on the measured device variation effect data in Fig. 2b-d. This resulted in only a modest accuracy decline for both models without any re-training or other mitigation techniques: the SSC accuracy dropped 2.39% to 82.01%, and the DVS-Gesture accuracy dropped 2.2% to 95.1%. Since the STM memristors need to implement the trained $\lambda$, we also analyzed the effects of variations in the device decay rate on model performance (Supplementary Figure 5 and Supplementary Note 3). These results show that the models exhibit negligible accuracy degradation with variations up to 10% of the mean, without any re-training or variation-aware training techniques. The noise and variation analyses support the resilience of the proposed model for practical analog hardware implementations, likely due to the use of only fully-connect layers and the shallow model architecture. Future implementation of techniques such as variation-aware-training[65] is expected to further improve the hardware accuracy.

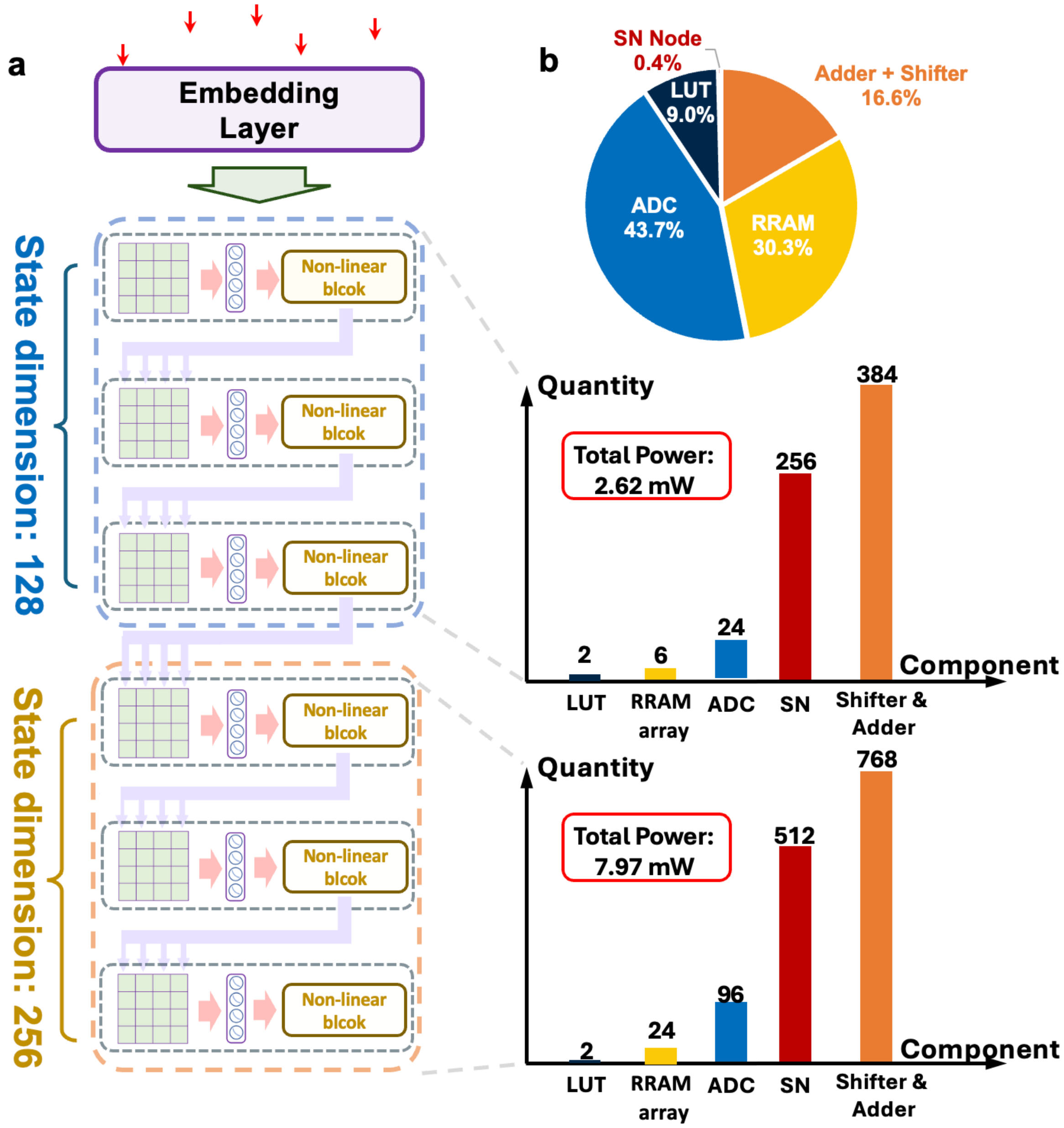

To estimate power consumption, we constructed simulation for the proposed hardware system for the SSC dataset, using experimentally measured data from the CIM chip and STM memristor devices for the SSC dataset. The system consists of 6 SSM blocks, as shown in Fig. 4a.

The total power consumption for our system is estimated to be 34 mW and the energy distribution across system components is shown in Fig. 4b (see **Method**). The 8-bit ADCs dominate the power budget, accounting for 43.7% of total power usage. Analog power consumption in the RRAM arrays follow, consuming 30.3%. SN devices for update behaviors consume 0.4%. However, we note that the estimate was based on ADCs and other digital components constantly operating at 40MHz due to the limitations of the current CIM chip. In practice, when the operations are truly operated in an event-driven fashion, the ADCs will not be active all the time and actual power consumption can be significantly lower. The power can be further improved by eliminating the analog-digital transitions to implement all operations in the analog domain, which is compatible with SSM based approaches.

# Discussion

We presented a co-design methodology for implementing SSMs on CIM hardware, targeting real-time, event-driven processing. A key principle is to enable asynchronous operations and perform all major functions including vector-matrix multiplication and state evolution natively in physics. This strategy yielded strong performance on diverse event-driven audio and vision datasets while achieved state-of-the-art accuracy and parameter efficiency compared to existing methods. We believe models such as SSMs, which offer excellent sequence modeling capability without requiring non-bio realistic operations including convolutions, provide a promising path towards realizing practical and energy-efficient neuromorphic systems. Our co-design philosophy, leveraging device and circuit physics for core computational primitives of SSMs, makes it particularly attractive for implementing these advanced sequence models on resource-constrained devices.

Unlike SNNs, which often leverage both input sparsity and internal activation sparsity (i.e., neurons only compute when they spike), our current SSM implementation does not enforce sparse internal activations. The system's efficiency is primarily derived from leveraging the input sparsity of the event-based data; active computation (e.g., VMM) is needed only when an event arrives, while the state evolution is handled passively by device physics. Inducing sparsity in the internal SSM activations, perhaps through thresholding or other mechanisms, could offer a path to even greater energy savings and remains a direction for future research.

We note the intrinsic short-term memory of physical devices like $WO_x$ has also been used in networks such as reservoir computing (RC)[42] to process temporal information. However, the proposed SSM model is fundamentally different from reservoir computing. In particular, the SSM model, including the state dynamic governed by the decay parameter $\mathbf{\Lambda}$, is trained end-to-end, whereas in RC systems only the output (or readout) layer needs to be trained and the internal reservoir is not trainable. Although the decay constants are fixed in the proposed SSM during inference to align with hardware physics, their values are not random. Instead, they are systematically determined by first training them as learnable parameters and then fixing them to a layer-wise average that proved effective for the task.

Our current implementation assumes static, non-input-dependent state transition matrix, a design choice mirroring early SSMs like S4 and S5 that prioritizes straightforward mapping onto weight-stationary CIM hardware. While highly efficient and friendly to the CIM architecture, this static approach prevents the model from dynamically adjusting its parameters based on the input content which is a key capability behind the high performance of modern selective SSMs such as Mamba. However, as recent device-level research demonstrates, the time constant of the SN can be actively modulated in-situ by an external control signal, such as a bias voltage[66]. Such approaches make it possible to dynamically adjust the state decay rates on-the-fly and effectively implement modern selective SSMs in the proposed architecture. Additionally, systematical shifts, such as temperature effects on the device's decay constant, can potentially be actively compensated by this bias voltage control, significantly enhancing the system's real-world robustness.

Since the SN state is updated when an event arrives, the endurance of the SN devices needs to be high enough as each update constitutes a write operation. It is important to note that these updates are small and incremental which is generally less stressful for the device, rather than full SET/RESET cycles. The $WO_x$ memristors used in our work have demonstrated programming endurance for more than $10^8$ cycles[67,68]. Additionally, other STM memristors such as those based on metal-insulator transition ($NbO_x$)[69], or spin-transfer torque devices[70], have shown even longer endurance and can potentially be used in the proposed architecture as well.

Finally, in this study we prioritized computational fidelity and robustness, including the use of differential positive and negative weight arrays and 1T1R cells in the CIM chip. More advanced techniques such as 2's complement[19] and passive crossbar arrays[71] can potentially also be used, offering benefits of higher compute density. Fully analog implementations that eliminate ADCs and DACs can also further improve the system's energy efficiency, achieving clockless asynchronous signal processing.

# Methods

Model Initialization

We first assemble a modified HiPPO–LegS[72] operator by computing a lower-triangular kernel whose $(i, j)$ entry (for $j \le i$) is given by the product of $\sqrt{1+2i}$ and $\sqrt{1+2j}$, with its diagonal entries shifted by subtracting the row index. To render the operator diagonalizable by an orthogonal transform, we augment it with a rank-one correction constructed from a vector of $\sqrt{i+1/2}$ terms and then enforce strict symmetry by averaging the result with its transpose. The resulting real symmetric matrix **S** admits an orthogonal eigen-decomposition

$$\mathbf{S} = \mathbf{V}\, diag(\mathbf{\Lambda})\mathbf{V}^T$$

in which $\mathbf{\Lambda}$ is guaranteed to consist solely of real values.

All subsequent projections—including the input mapping B and any output weights—are initialized using conventional real-valued schemes (e.g., LeCun normal[73]) without splitting into real and imaginary parts. By maintaining this wholly real-valued pipeline—from operator construction through eigen-decomposition and parameter initialization—our method produces an SSM block whose state updates and nonlinear projections can be implemented directly on analog crossbar arrays and digital lookup tables, avoiding any reliance on complex hardware units.

Device Fabrication

The CIM chip consists of four 64 × 64 crossbar arrays based on 1T1R cells with a physical diameter of 200 nm[75]. These crossbars were monolithically integrated above CMOS circuitry using a 65 nm process. Each tile incorporates 8-bit DACs for input encoding and shared 8-bit ADCs for output digitization. A differential (dual-column) scheme is employed to represent signed weights, enabling symmetric mapping and improved linearity. The peripheral circuits and control logic are managed by an on-chip RISC-V processor to enable fully integrated and autonomous CIM operation. The CIM chip is measured using a custom-designed test board.

The $WO_x$ memristor devices with a size of 1μm × 1μm were fabricated on $SiO_2$/Si substrate, following the methodology outlined in our previous work[41,44]. The bottom electrode (BE) patterns were defined by optical photolithography (GCA AS200 AutoStep), followed by 60 nm W metal deposition by DC reactive sputtering using a W metal target at room temperature, and subsequently formed using a lift-off process. A 250 nm thick $SiO_2$ film was deposited by plasma-enhanced chemical vapor deposition (PECVD), followed by directional etching using RIE (LAM 9400) to expose the BE top surface and create a spacer structure. Rapid thermal annealing (RTA) was performed to form the $WO_x$ layer, by oxidizing the exposed part of the W BE in $O_2$ atmosphere at 400 °C for 20-80 seconds. Afterwards, the Au/Ti top electrodes (TEs) were formed by optical lithography, liftoff and E-beam metal deposition. Additional RIE process was applied to select etch out the $WO_x$ expose the BEs for electrical contacts. The devices were characterized using a Keithley 4200S semiconductor analyzer with 4225-RPM module.

Parallelized BPTT

We train the model using backpropagation through time, computing gradients with respect to all temporal dependencies via automatic differentiation. The forward pass implements the discrete-time SSM recurrence $\mathbf{h}(t + \Delta t) = \overline{\mathbf{\Lambda}}(\Delta t)\mathbf{h}(t) + \overline{\mathbf{B}}\mathbf{x}(t + \Delta t)$ for sequences of length $L$. Rather than computing this recurrence sequentially, we leverage a parallel associative scan algorithm that exploits the associativity of linear recurrences through a operator $(\mathbf{A}_i, \mathbf{b}_i) \circ (\mathbf{A}_j, \mathbf{b}_j) = (\mathbf{A}_j \cdot \mathbf{A}_i, \mathbf{A}_j \cdot \mathbf{b}_i + \mathbf{b}_j)$. This formulation enables simultaneous computation of all hidden states via a balanced binary tree reduction, achieving $O(\log L)$ parallel time complexity compared to $O(L)$ for sequential evaluation given sufficient parallel processors[74]. During the backward pass, automatic differentiation through the scan operation computes gradients for all SSM parameters—including state matrix $\mathbf{\Lambda}$, input projection $\mathbf{B}$, output projection $\mathbf{C}$, and learnable discretization timesteps—while preserving the same logarithmic complexity. This parallelized BPTT implementation allows efficient gradient-based optimization on long event sequences while maintaining full temporal credit assignment across all time steps.

Theoretical FLOPs Analysis

We conducted a theoretical FLOPs analysis to compare the computational complexity of Event-SSM and ResNet models for DVS-Gesture data processing. The Event-SSM model has 2 stages each composing of 3 SSM blocks. The state in the first stage has the dimension of 128 while the second has 256. We assume the input has sequences of 65,536 events which is typical to DVS-Gesture dataset.

For models based on synchronous processing, event data are converted to 30fps video (approximately 180 frames for 6 seconds), with each frame having a resolution of 128×128×2 channels (ON/OFF events). We analyzed both ResNet-18 and ResNet-50 as standard architectures.

Our analysis assumes: (1) multiply-accumulate (MAC) operations count as 2 FLOPs, (2) matrix multiplication $H_{\text{in}} \times H \times H_{\text{out}}$ counts as $2 \times H_{\text{in}} \times H \times H_{\text{out}}$ FLOPs, and (3) normalization operations count as 5 FLOPs per element. (4) ResNet processes each frame independently.

For Event-SSM, we calculate FLOPs for each layer including embedding projection $(L \times D \times 2)$, and for each SSM layer: input projection $(L \times H_{\text{in}} \times H \times 2)$, state evolution $(L \times H_{\text{in}} \times 2)$, output projection $(L \times H \times H_{\text{out}} \times 2)$, feedforward network $(L \times H_{\text{out}}^2 \times 2)$, and normalization $(L \times H \times 5)$, where $D$, $L$, $H_{\text{in}}$, $H$, $H_{\text{out}}$ denote embedding dimension, sequence length, input feature dimension, and state dimension, output feature dimension respectively. Pooling with stride 16 reduces the sequence length from 65,536 to 4,096 after the first stage, and to 256 after the second stage. For ResNet, we use the standard formula $FLOPs = H_{\text{out}} \times W_{\text{out}} \times C_{\text{out}} \times (K^2 \times C_{\text{in}} \times 2)$ for each convolutional layer, where $H_{\text{out}}$, $W_{\text{out}}$, $C_{\text{out}}$, $K$, $C_{\text{in}}$ denote the output height, output width, number of output channels, kernel size, and number of input channels, respectively. Total FLOPs equal per-frame FLOPs multiplied by the number of frames (180)

Results show that Event-SSM requires 1.68 GFLOPs, while ResNet-18 and ResNet-50 require 104.28 and 219.80 GFLOPs, respectively. This corresponds to $62.12\times$ and $130.93\times$ computational efficiency improvements for Event-SSM, primarily attributed to (1) the sparsity of event representation (65,536 events vs. 5,898,240 pixel values, a $90\times$ density difference), (2) elimination of fixed frame-rate processing overhead, (3) compact model design, and (4) real-valued operations throughout.

Noise Modelling

To evaluate the model's robustness against hardware non-idealities, we conducted a noise-injection simulation. In this simulation, we injected noise after each computational step, namely VMM and the state update in the SNs, to the physical hardware. The injected noise was modeled using a Gaussian distribution ($\mathcal{N}\ (\mu\ ,\sigma^2)$) based on values empirically derived directly from the experimental characterization data of the fabricated CIM chip and SN devices. This methodology ensures that our simulation reflects the analog noise profile inherent in the physical system.

Power Analysis

The power for the CIM chip, which performs the VMM, was measured directly on the PCB board. We measured the power consumption while the chip was performing a typical VMM operation at its operational clock frequency of 40 MHz. Energy consumption is calculated as Power Consumption per array × number of arrays needed and the measured power consumption per array is about 0.028mW. It is important to note that, due to the limitations of the current test chip, this measurement was conducted under a global clock, and the clock frequency is much higher than the requirement of the dataset (After pooling, SSC dataset will have roughly 5000 events/second, corresponding to operation at the kilohertz level). We therefore believe that a future implementation using a fully event-driven CIM chip would exhibit even lower power consumption, as it would theoretically only perform VMM computations when an input event arrives, thus fully leveraging data sparsity.

Power consumption of the STM memristor array is dynamic and event-driven. Therefore, we estimated the total energy $E_{\mathrm{SN}}$ by summing the energy of each individual inputs. We used the methodology from Yoo et al.[76], as shown in the following equation:

$$E_{\mathrm{SN}} = \sum_{n=1}^{N} V_{\mathrm{pulse}}^2 * G\big(x_n, \gamma_{n,} p_n\big) * t_{\mathrm{pulse}} \tag{10}$$

where $G\big(x_n, \gamma_{n,} p_n\big)$ represents the conductance of the specific SN device being addressed. For each SN updates, which corresponds to an input spike x(t+Δt) in Eq. 8, a burst of *N* 1.4 V per 50 μs pulses were applied on the memristor, where the number of pulses *N* is proportional to the VMM output.

Data from the remaining components are obtained from synthesized circuits using TSMC 28nm node, following an approach by Wang et al.[24] Inside each block, positive and negative weight values are mapped onto separate RRAM arrays. The 8-bit input activations are applied in a bit-serial fashion using two voltage pulses, each representing a 4-bit value (16 discrete levels). The resulting analog currents from RRAM arrays are digitized by 8-bit ADCs and subsequently shift-added digitally to obtain an 8-bit partial sum, representing the output activation values that are applied to the corresponding SN devices. In our model, the layer at the front uses a smaller state dimension and the layers at the back use a lager state dimension to allow for richer information compressed across time. Each SSM block also require two LUTs to implement the non-linear functions, namely, GELU and Sigmoid. The power of these nonlinear functions are similarly simulated. Given that these functions operate on the 8-bit quantized values from ADCs, each non-linear operation can be efficiently realized using a compact LUT containing only $2^8$ (256) pre-computed entries. The power for one LUT is estimated to be 0.25 mW at 40 MHz.

## Data Availability

All data presented in this study are either included in this article and its Supplemental Information or are available upon request to the corresponding author. Source data are provided with this paper.

## Acknowledgements

This work was supported in part by the National Science Foundation under Grant CCF-2413293 and by the Michigan Semiconductor Talent and Technologies for Automotive Research program. W. D. L acknowledges support from the James R. Mellor Professorship. We also acknowledge technical support from Lurie Nanofabrication Facility.

## Author contributions

W.D.L conceived the project. X.Z. and M.H. designed the experiments. X.Z., S.L. and E.Y.L performed model training. M.H. and S.K. fabricated devices and performed measurements. X.Z. and Y.L performed circuit simulations. All the authors contributed to the discussion and contributed to the manuscript writing at all stages.

## Competing interests

The authors declare no competing interests.

# Tables

| Method | Spiking Heidelberg Digits | | | Spiking Speech Commands | | |
|---|---|---|---|---|---|---|
| | **Test Acc.** | **Params** | **Async.** | **Test Acc.** | **Params** | **Async.** |
| Hammouamri et al.[47] | 95.1% | 0.2M | ✗ | 80.7% | 2.5M | ✗ |
| Bittar and Garner[48] | 94.6% | 3.9M | ✗ | 77.4% | 3.9M | ✗ |
| Sun et al.[49] | 92.5% | 0.1M | ✗ | – | – | – |
| Cramer et al.[5] | 92.4% | – | ✗ | – | – | – |
| Dampfhoffer et al.[50] | – | – | – | 77.0% | – | ✗ |
| **Ours** | **95.7%** | **0.3M** | ✓ | **84.7%** | **0.6M** | ✓ |

**Table 1 |** Comparison of our SSM model to the state-of-the-art SNN based models on Spike Audio Datasets in simulation

| Method | DVS128 Gesture | | | DVS128 Lips | | |
|---|---|---|---|---|---|---|
| | **Test Acc.** | **Params** | **Async.** | **Test Acc.** | **Params** | **Async.** |
| **SNN-based methods** | | | | | | |
| She et al.[51] | 98.0% | 1.1M | ✗ | – | – | – |
| Apolinario et al.[52] | 97.7% | 1.6M | ✗ | – | – | – |
| Wang et al.[53] | 97.1% | 1.5M | ✗ | 42.2% | 8.1M | ✗ |
| Bulzomi et al.[54] | – | – | – | 60.2% | 47.0M | ✗ |
| **RNN-based methods** | | | | | | |
| Innocenti et al.[55] | 97.7% | – | ✗ | – | – | – |
| Subramoney et al.[56] | 97.8% | 4.8M | ✗ | – | – | – |
| **CNN-based methods** | | | | | | |
| Tsourounis et al.[57] | – | – | – | 63.2% | 40.5 M | ✗ |
| Wang and Zhao et al.[58] | – | – | – | 34.5% | 64.6 M | ✗ |
| **Tan et al.[4]** | – | – | – | **72.1%** | **38.5 M** | ✗ |
| **Other methods** | | | | | | |
| Martin-Turrero et al.[59] | 94.1% | 14 M | ✓ | – | – | – |
| Peng et al.[60] | 97.9% | 4.5 M | ✗ | 69.8% | – | ✗ |
| Gehrig et al.[61] | – | – | – | 48.7% | 21.5 M | ✗ |
| **Liu et al.[62]** | **98.8%** | – | ✗ | – | – | – |
| **SSM-based methods** | | | | | | |
| Schöne et al.[36] | 97.7% | 5 M | ✓ | – | – | – |
| **Ours** | **97.3%** | **5 M** | ✓ | **63.5%** | **5.7 M** | ✓ |

**Table 2 |** Comparison of our SSM model to other leading models on the DVS128 Gesture and DVS128 Lips datasets in simulation.

# Figure Legends/Captions

**Fig. 1 | Model architecture for the proposed event-driven State Space Model (SSM). a** High-level network architecture, including a channel-wise embedding layer (green) that maps incoming events to a high-dimensional latent state space and stacked SSM blocks (red) that asynchronously process information in the state space; **b** Functional diagram of the channel-wise embedding layer. An incoming spike activates the embedding layer which produces a D-dimensional vector depending on the event's channel index. **c** Operational schematic of a single SSM block (upper). The state of the SSM is represented by those of individual state nodes (SNs) (middle) in the proposed implementation. The state in an individual SN decays exponentially following a trained decay time constant. When an input **x** arrives, it is projected onto the state space through $\overline{\mathbf{B}}\mathbf{x}$ and is added to the current state. The updated state **h** is then read out, passed through some non-linear functions and sent to the next block. Note actions (i.e. computations) are only performed when an event arrives, allowing asynchronous operations.

**Fig. 2 | Hardware implementation of an SSM block. a** Schematic of the proposed hardware implementation. The state update function, shown at the bottom, is physically realized through an Resistive RAM (RRAM) crossbar array that performs vector-matrix multiplication (VMM) operation between the $\overline{\mathbf{B}}$ matrix (blue) and the input vector (yellow), and an array of state nodes (SNs) that stores the state and perform the diagonalized state evolution functions (red). These functions directly map the correspondingly colored operations in the state update function. **b** VMM operations between the $\overline{\mathbf{B}}$ matrix and the input vector implemented in an RRAM compute-in-memory (CIM) chip. The experimentally measured VMM outputs (Analog-to-Digital Converter (ADC) Code) from the RRAM CIM chip (dots) are plotted against ideal values (line). Upper inset: photo of the CIM chip. Lower inset: deviation of the measured data from the ideal VMM values due to device variations. A standard deviation of 4.6 least significant bit (LSB) was measured. **c-d** State evolutions implemented in $WO_x$ memristors with short-term memory, showing experimentally measured memristor conductance evolutions (dots) vs. expectations obtained from Eq. 8 (lines), for two required decay rates of 0.35 ms-1 c and 0.2 ms-1 d. Each memristor act as an SN whose conductance represents a component of the state and performs state evolutions. The inset in d shows a scanning electron microscope (SEM) image of the fabricated $WO_x$ memristor array. (scale bar: 20 µm).

**Fig. 3 | Effect of model depth and decay parameter ($\lambda$) configuration on classification accuracy. a-d** Impact of the number of sequential SSM blocks on accuracy across different event-driven datasets: Spiking Heidelberg Digits (SHD), Spiking Speech Commands (SSC), and Dynamic Vision Sensor (DVS) datasets (DVS128 Gesture and DVS128 Lips). **e** Comparison of classification accuracy with fixed (yellow) vs freely learned (blue) decay parameter $\lambda$. **f** Layer-wise analysis of the decay constant ($\tau = 1/\lambda$) for different datasets, showing large t exist for the SSC dataset which can lead to instability during training.

**Fig. 4 | Hardware architecture and power distribution for the SSC dataset implementation. a** Architecture of the hardware implementation for the SSC dataset. Arrows represent data flow within an SSM block and between SSM blocks. The accompanying bar charts list the quantity of hardware components for each stage. Components include Look-Up Table (LUT), RRAM array, ADC, SN and Shifter & Adder. **b** Power distribution across key hardware components for the SSC dataset implementation.

# Supplementary Information

## Supplementary Figures

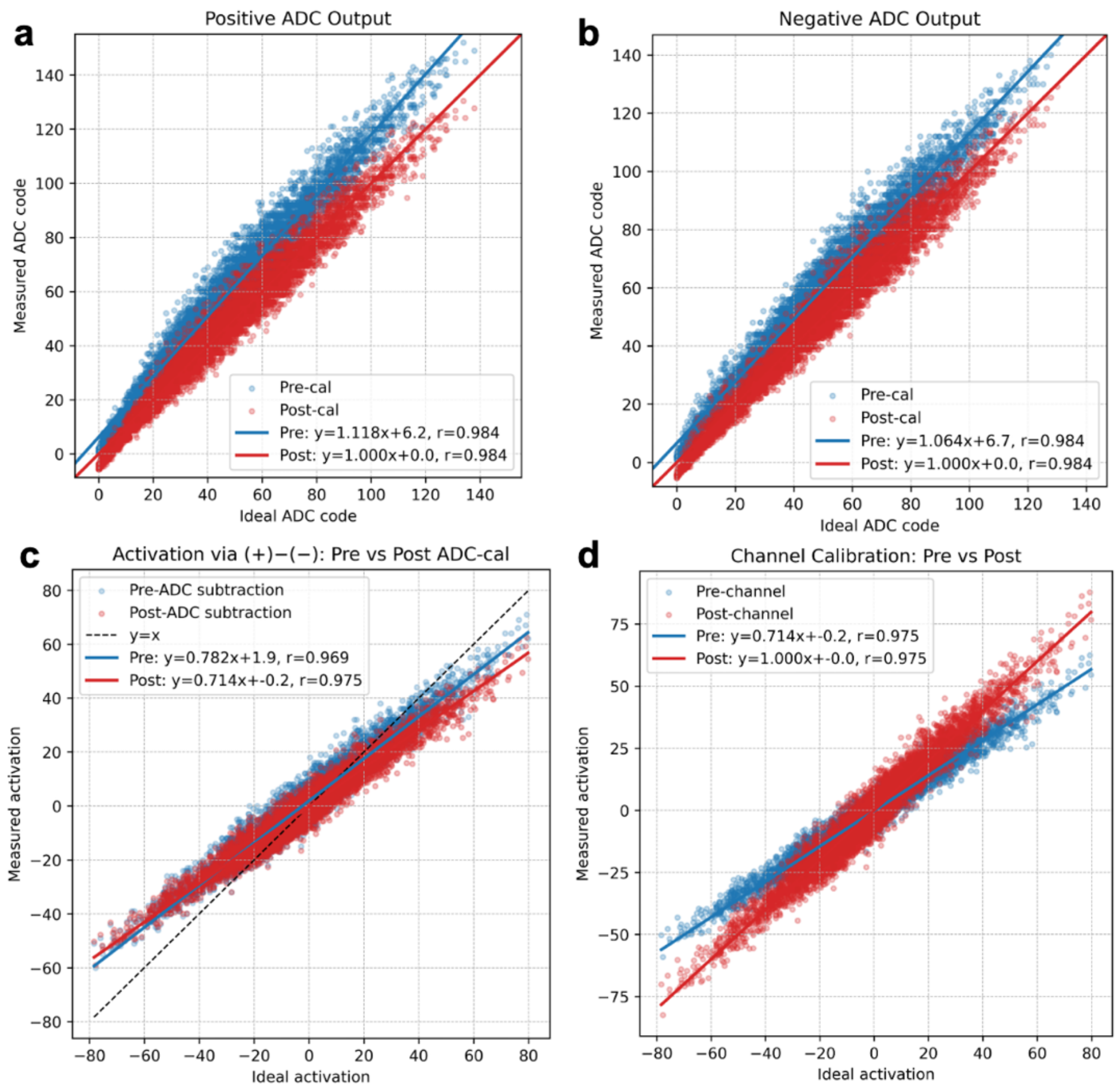

**Supplementary Figure 1**. Minimizing systematic errors through a two-stage calibration process. **a, b** Pre-calibration (blue) and post-calibration (red) outputs for the positive and negative columns through ADC, respectively, demonstrating the correction at the individual ADC-level (Stage I). **c** The combined activation after Stage I (ADC-level calibration and subtraction). While the correlation is much improved, some residual deviation from the ideal identity line (dashed black) remains. **d** The final output after Stage II (channel level calibration), which aligns almost perfectly with the ideal y=x relationship.

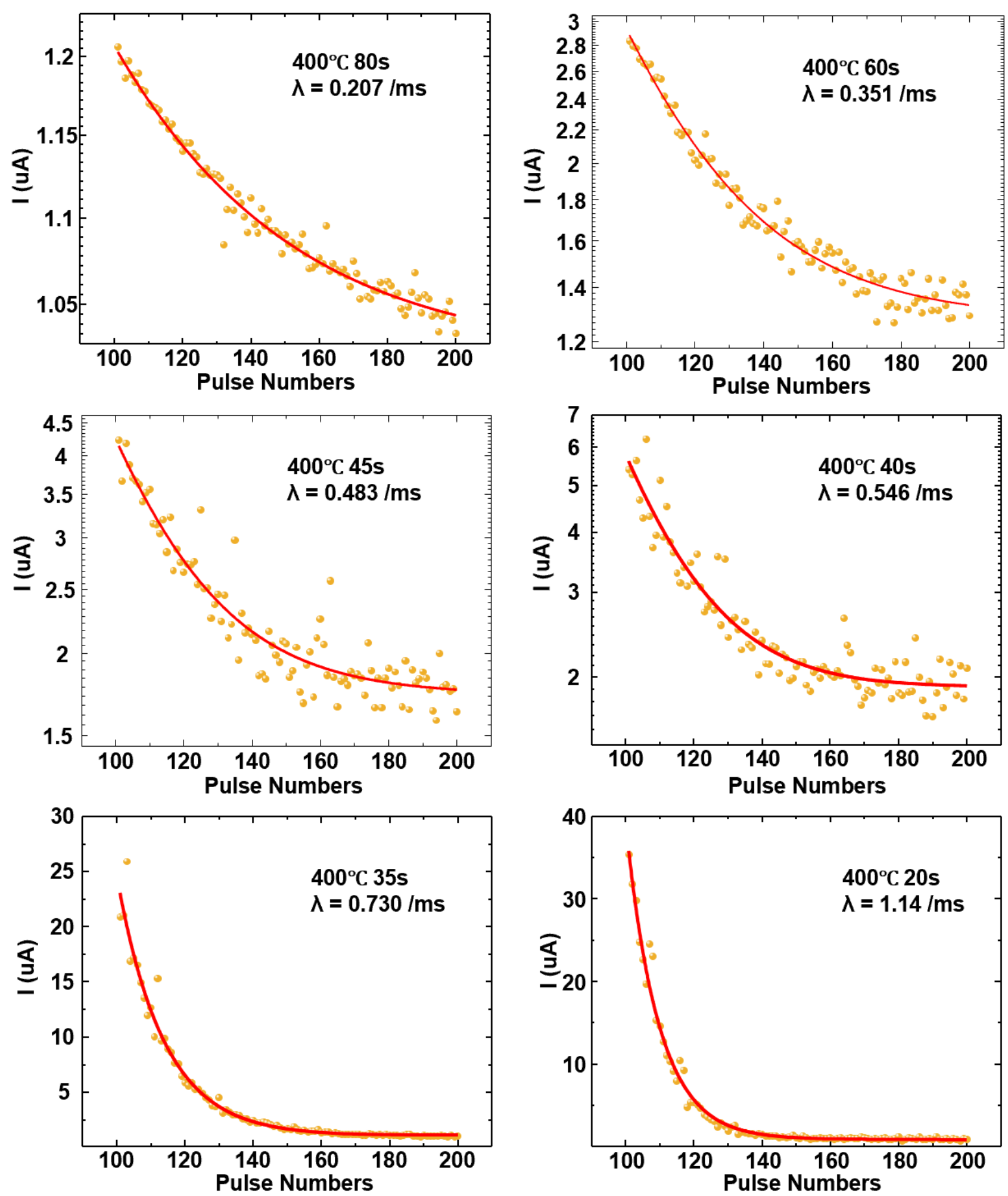

**Supplementary Figure 2.** Fabricated STM memristors with different decay rates The STM memristor's decay behavior can be tuned during the rapid thermal processing (RTP) process. The decay rate (λ) can be effectively tailored by varying the RTP time while keeping the annealing temperature at 400 °C. This effect is attributed to changes in the oxygen vacancy density within the oxide layer: longer RTP durations reduce vacancy density, resulting in slower decay rates, as reflected in smaller λ values. The extracted decay constants span a broad range ($\lambda = 0.207\ ms^{-1}$ to $1.14\ ms^{-1}$), confirming that RTP duration serves as a viable process-level knob for engineering memristor temporal dynamics. The λ values were obtained by fitting the experimental data (yellow dots) to an exponential function (red curves).

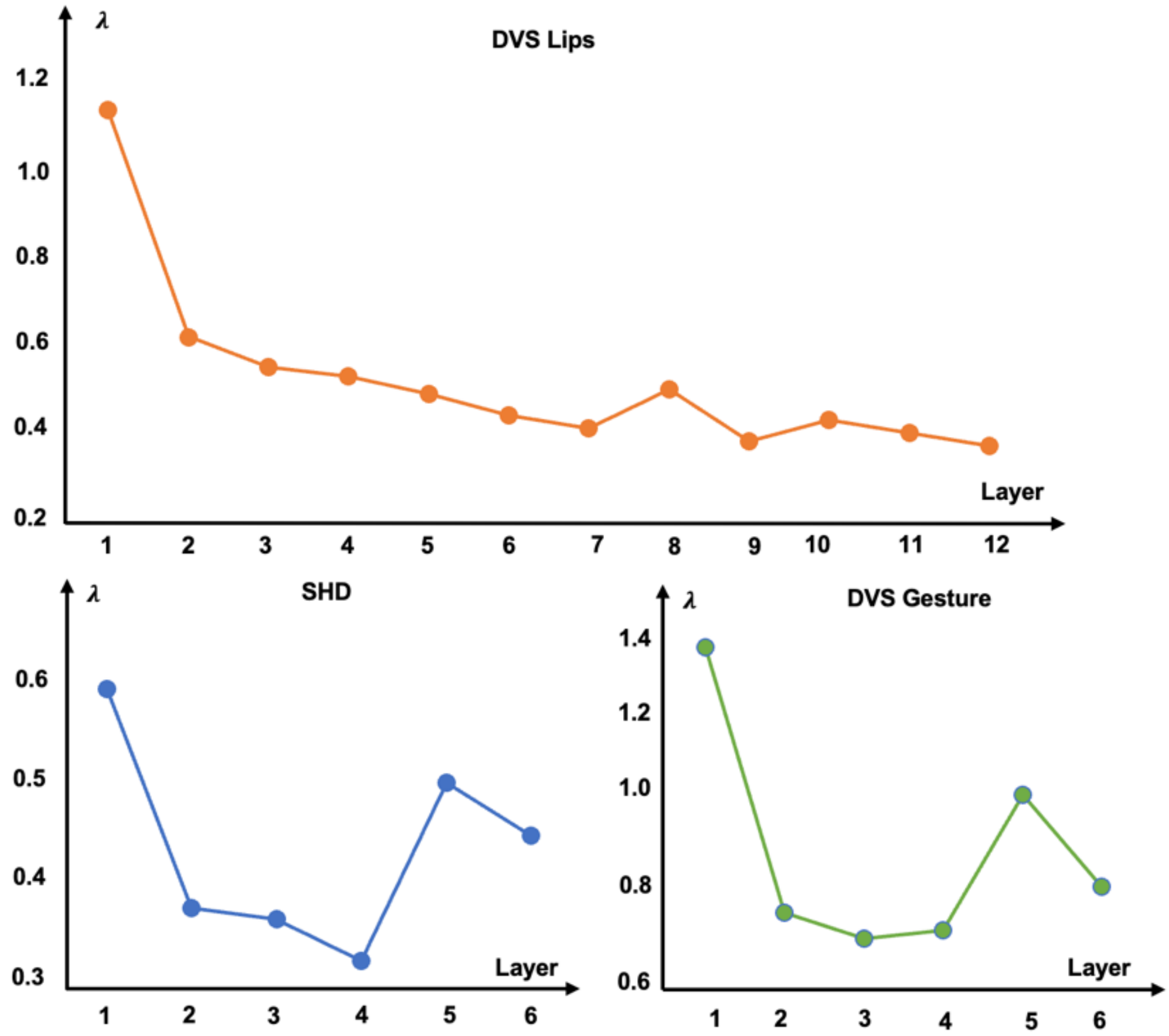

**Supplementary Figure 3.** Learned $\lambda$ values across layers when trained on DVS Lips, DVS Gesture and SHD dataset

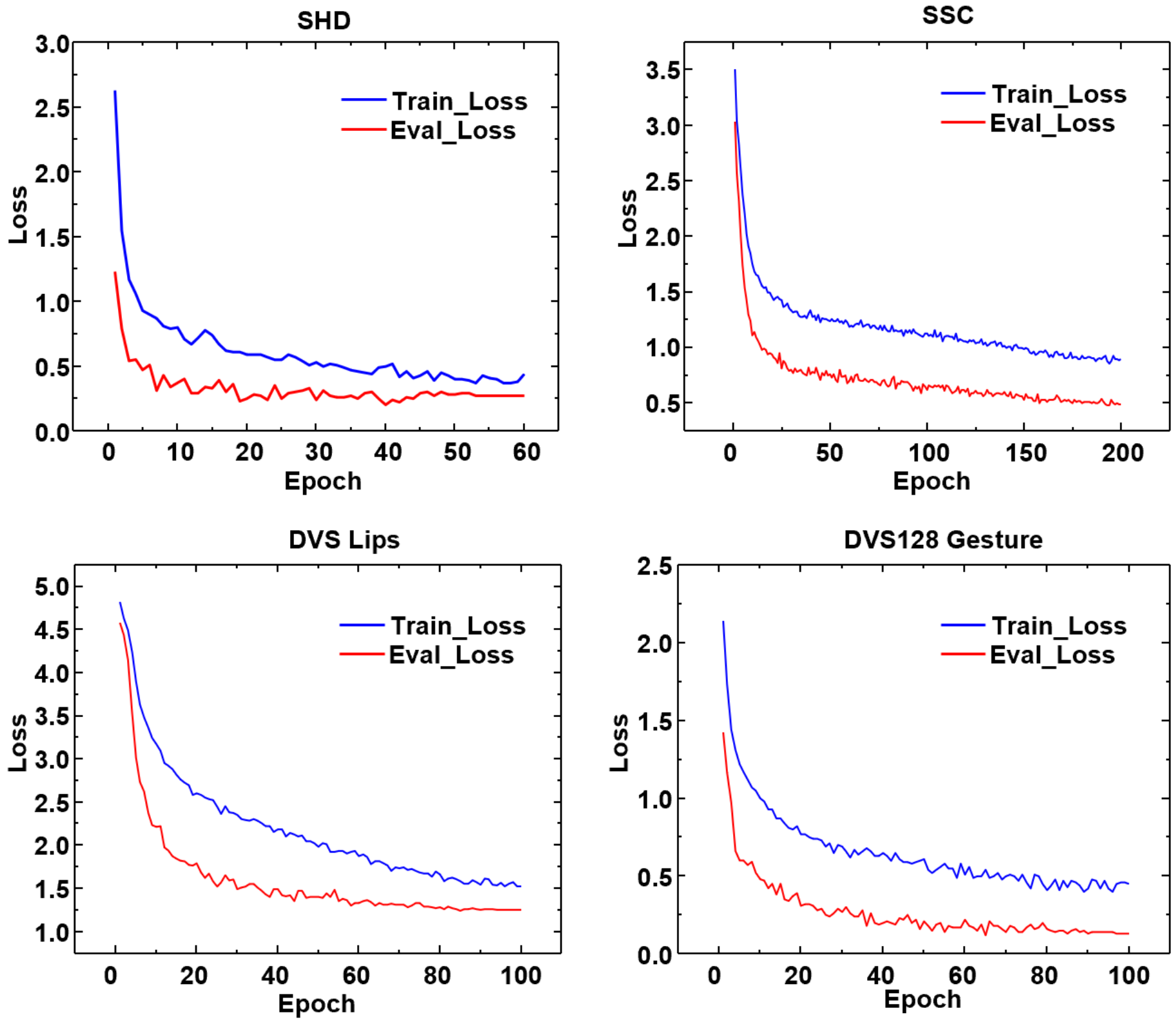

**Supplementary Figure 4.** Training loss (blue line) and evaluation loss (red line) versus training epochs for the model for the four event-driven datasets (Spiking audio tasks SHD and SSC, and the event-based vision tasks DVS Lips and DVS128 Gesture).

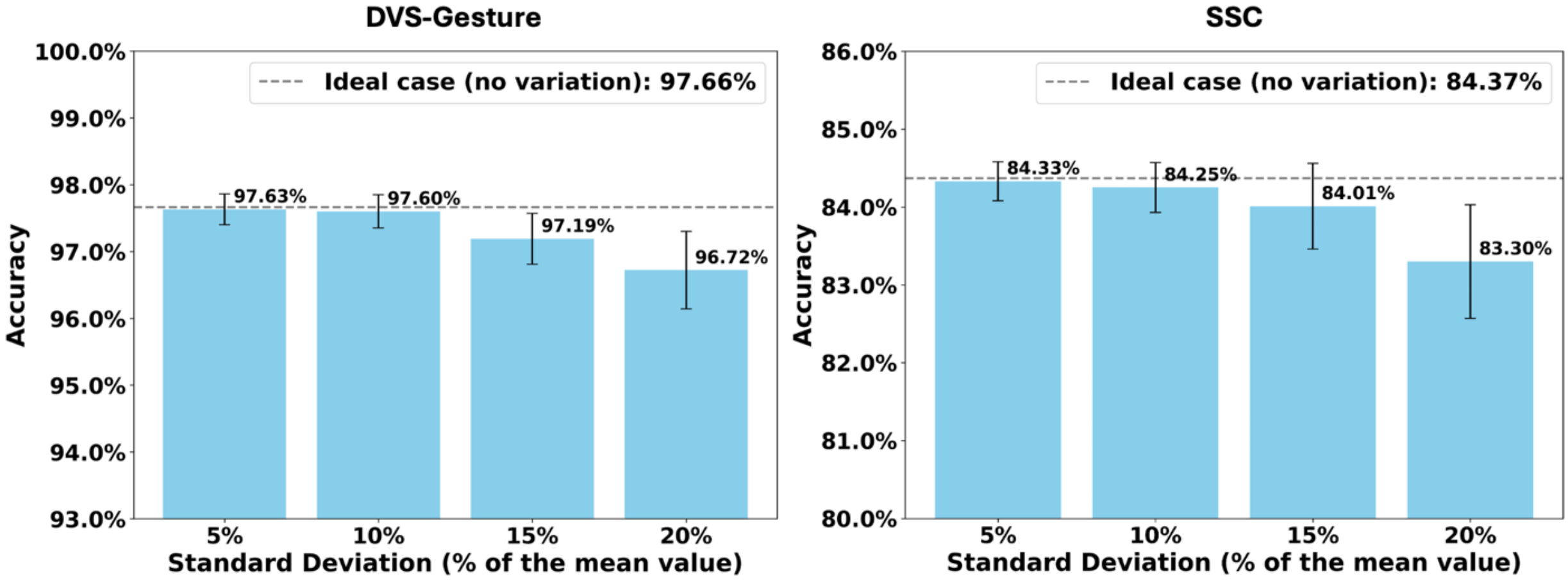

**Supplementary Figure 5.** Model accuracy as a function of device variations in the decay constant λ. The decay constant λ of the STM memristor devices are modeled to follow a Gaussian distribution centered on the target mean value, and the standard deviation of the distribution is increased from 5% to 20% of the mean. Error bars represent results from 40 independent test runs at each variation level. The dotted line represents the ideal baseline accuracy without variation from the quantized-weight model.

# Supplementary Tables

| Dataset / $\lambda(ms^{-1})$ | SHD | SSC | DVS-Gesture | DVS-lips |
| --- | --- | --- | --- | --- |
| First layer | 0.55 | 0.35 | 1.15 | 1.15 |
| Other layers | 0.35 | 0.20 | 0.70 | 0.50 |

**Supplementary Table 1** $\lambda$s selected for each dataset.

# Supplementary Notes

### Supplementary Notes 1: Selection of Fixed $\lambda$ Values

To determine the fixed $\lambda$ values, we examine $\lambda$ learned from unconstrained models on various event-driven datasets. Notably, the first SSM layer tends to converge to a distinctly larger $\lambda$ than subsequent layers, reflecting its role in rapidly filtering or dampening high-frequency input components. Deeper layers, by contrast, stabilize around smaller $\lambda$ values that favor the accumulation of longer-term information. A special case is the SSC dataset, where unconstrained $\lambda$ updates lead to divergence before the model fully converges. In this scenario, we extract $\lambda$ values from the last stable iteration before encountering numerical errors. Based on these observations, our fixed-$\lambda$ design adopts a two-tier approach: For the first layer, we assign a larger $\lambda$, mirroring the heightened filtering observed in the unconstrained models. This ensures the network can still capture the fast-decay dynamics critical for early-stage feature extraction. Then, we use a single shared $\lambda$ for all remaining layers. Empirically, we set this to the approximate average of the layer-wise $\lambda$s found in the unconstrained model (excluding the first layer). The training then continues with the fixed $\lambda$s until final convergence.

## Supplementary Notes 2: Calibrating the output of RRAM CIM chip

The analog nature of compute-in-memory (CIM) systems introduces several sources of error. Non-idealities in the RRAM devices and readout circuitry lead to a discrepancy between the ideal and measured VMM outputs. To address these cumulative errors, we employ a two-stage linear calibration procedure: an ADC-level calibration followed by a channel-level activation calibration.

ADC-level Calibration: The first stage corrects errors at the individual ADC. Since our system uses a differential mapping scheme, where positive ($W^+$) and negative ($W^-$) components of a weight are mapped to separate physical columns, the corresponding positive and negative ADC channels are calibrated independently. For each column $c$ (either positive or negative), we perform a linear regression to model the relationship between the measured raw ADC codes ($y_c$) and the ideal codes ($x_c$), such that $y_c \approx s_c x_c + b_c$. The transformation is then applied to correct the raw output, yielding the ADC-calibrated code $\widehat{y_c}$: $\widehat{y_c} = a_c y_c + d_c$ where $a_c = \frac{1}{s_c}$ and $d_c = -\frac{b_c}{s_c}$. This step aligns the output of each physical channel to its ideal value as shown in Supplementary Figure 1a and b. The ADC-calibrated activation is then obtained by subtracting the calibrated negative column 's output from the calibrated positive column's output (Supplementary Figure 1c).

Channel-level Calibration: The second stage corrects remaining channel-level (the difference of the calibrated positive column and the corresponding negative column) errors. We regress the ADC-calibrated channel activation ($\widehat{y_k}$) for a given channel $k$ on the final ideal activation ($x_k$), modeling the relationship as $\widehat{y_k} \approx \alpha_k x_k + \beta_k$. Following the same principle, we get calibrated output $\widetilde{y_k} = A_k \widehat{y_k} + B_k$ where $A_k = \frac{1}{\alpha_k}$ and $B_k = -\frac{\beta_k}{\alpha_k}$.

These two sequential linear corrections can be mathematically fused into a single operation for direct physical hardware calibration. By substituting the Stage 1 equations into the Stage 2 equation, we derive a single transformation from the raw ADC codes ($y_{c,pos}$, $y_{c,neg}$) to the final output ($\widehat{y_k}$):

$$\widetilde{y_k} = A_k \widehat{y_k} + B_k = A_k \left(\widehat{y_{c,pos}} - \widehat{y_{c,neg}}\right) + B_k = A_k \left[\left(a_{c,pos}\, y_{c,pos} + d_{c,pos}\right) + \left(a_{c,neg}\, y_{c,neg} + d_{c,neg}\right)\right] + B_k = (A_k\, a_{c,pos})\, y_{c,pos} - (A_k\, a_{c,neg})\, y_{c,neg} + [A_k(d_{c,pos} - d_{c,neg}) + B_k]$$

This fused equation provides the exact target parameters for physically tuning the ADC hardware: The $W^+$ ADC's gain is set to $G_{pos} = A_k\, a_{c,pos}$, and the $W^-$ ADC's gain is set to $G_{neg} = A_k\, a_{c,neg}$ and The total differential offset is set to $O_{diff} = A_k(d_{c,pos} - d_{c,neg}) + B_k$. In practice, this allows us to program these fused coefficients directly into the circuitry (e.g., by adjusting ADC reference voltages for gain and comparator biases for offset). This corrects all linear errors at the point of conversion, achieving the final calibrated result shown in Supplementary Figure 1d and Fig. 2b without the overhead of sequential digital post-processing.

**Supplementary Notes 3: Model's Performance against Decay Constant ($\lambda$) Variations.**

Physical device fabrication is subject to inherent process variability, meaning the actual decay constant $\lambda$ of a fabricated STM memristor device will likely deviate from its intended target value. To analyze the model's sensitivity to device-to-device variations, we modeled the device decay constant $\lambda$ as following a Gaussian distribution centered on the layer's target mean $\lambda$. The standard deviation of the distribution was increased from 5% to 20% of the target mean $\lambda$. The model's accuracy was then evaluated by randomly assigning $\lambda$ values to the STM memristor devices following the distribution, and repeated 40 times for each distribution to obtain statistically significant results.

The results, plotted in Supplementary Figure 5, show that the model maintains high accuracy even with substantial $\lambda$ variations up to 10% of the target mean $\lambda$. This result suggests that the proposed network can tolerate inherent variability of physical devices, confirming the feasibility of our hardware co-design approach.